\documentclass[aps,prb,showpacs,preprint]{revtex4}
\usepackage{graphicx}

\def\etal{\emph{et al. }}

\begin{document}

\title{\begin{flushleft}Novel pinning phenomena in a superconducting film with a square lattice of artificial pinning centers\end{flushleft}}
\author{\begin{flushleft}Z. Jiang, D.A. Dikin\renewcommand{\thefootnote}{\alph{footnote}}\footnote[1]{$^)$Current address: Department of Mechanical Engineering, Northwestern University}$^)$, and V. Chandrasekhar\footnote[2]{$^)$Electronic mail: v-chandrasekhar@northwestern.edu}$^)$\vspace{-0.43cm}\end{flushleft}}
\affiliation{\begin{flushleft}\text{Department of Physics and Astronomy, Northwestern University, Evanston, IL 60208}\vspace{-2cm}\end{flushleft}}
\author{\begin{flushleft}V.V. Metlushko\vspace{-1.2cm}\end{flushleft}}
\affiliation{\begin{flushleft}\text{Department of Electrical Engineering and Computer Science, University of Illinois-Chicago, IL 60607}\vspace{-2cm}\end{flushleft}}
\author{\begin{flushleft}V.V. Moshchalkov\vspace{-0.43cm}\end{flushleft}}
\affiliation{\begin{flushleft}\text{Department of Physics and Astronomy, Katholieke Universiteit, Leuven, Belgium}\end{flushleft}}
%\date{\today}
\pacs{74.25.Qt, 74.25.Ha, 74.78.Db}

\begin{abstract}
We study the transport properties of a superconducting Nb film with a square lattice of artificial pinning centers (APCs) as a function of dc current, at a temperature close to the superconducting transition temperature of the film. We find that, at low dc currents, the differential resistance of the film shows the standard matching field anomaly, that is, the differential resistance has a local minimum at magnetic fields corresponding to an integer number of flux lines per APC. However, at higher dc currents, the differential resistance at each matching field turns to a local maximum, which is exactly opposite to the low current behavior. This novel effect might indicate that the flux lines in the APC system change their flow mode as the dc current is increased.
\end{abstract}

\maketitle

Superconducting films with a lattice of artificial pinning centers (APCs) have been investigated for several years \cite{fiory,victor1,vitali,pannetier,reichhardt1}. It is found that the vortex lattice is highly ordered at certain matching fields, $H_n$, where the number of flux lines equals an integer multiple $n$ of the number of APCs. This ordering effect provides additional pinning interactions that result in a local minimum resistance and maximum critical current at the matching fields. Interest has been growing in such systems in the presence of driving forces \cite{reichhardt2}. Although much work has been done in systems with random disorder \cite{vinokur,giamachi}, the behavior of APC systems is only beginning to be explored. In this letter, we investigate the transport properties of a superconducting Nb film with a square lattice of APCs as a function of dc current through the film. Our results show novel behavior in a specific range of the dc current, which has not been observed in previous experiments.

Figure 1(a) shows a schematic of the device discussed in this letter. This device was patterned onto an oxidized Si substrate using laser interferometric lithography. The entire device is made of a single layer of Nb film (100 nm thick) with a square lattice of APCs with period $d$$\approx$1 $\mu$m and diameter $D$$\approx$0.3 $\mu$m. A scanning electron microscopy (SEM) image of the Nb film is shown in Fig. 1(b) with a 2 $\mu$m size bar. Technical details of the patterning procedure can be found in Ref. [3].

Figure 2 shows the differential resistance of the Nb film as a function of magnetic field normal to the film with different dc bias currents ($I_{dc}$) through the sample. The resistance was measured by conventional four-terminal techniques using an ac resistance bridge. The temperature of the system was fixed at $T$$=$7.07 K, close to the superconducting transition temperature, $T_c$$=$7.13 K, of the film.  All the resistance curves show oscillations as a function of magnetic field, with a fundamental period of $\simeq$ 18 G, corresponding to one superconducting flux quantum $h/2e$ per unit cell of the square lattice, or alternately, one flux line per APC.  At low dc bias currents ($I_{dc}$$\leq$100 $\mu$A), the differential resistance curve has a local minimum at magnetic fields $H_n$ corresponding to an integer number $n$ of flux lines per APC.  This behavior is the one that is conventionally observed. It is associated with strong pinning of flux lines at the matching fields $H_n$, leading to lower dissipation associated with flux line motion, and hence a lower resistance.  At higher dc bias currents ($I_{dc}$$\geq$110 $\mu$A), however, the local minimum becomes a maximum at each matching field, exactly opposite to the results at the low dc current. The current at which this change occurs is between 100 $\mu$A and 110 $\mu$A, where the behavior of the differential magnetoresistance shows a dramatic change in a small current range of 10 $\mu$A.  At even higher dc bias currents ($I_{dc}$$>$160 $\mu$A), not plotted out in Fig. 2, the magnetoresistance oscillations are washed out.

The novel behavior observed in Fig. 2 can also be seen in Fig. 3(a), the differential resistance measurements as a function of dc current at two different magnetic fields of 18 G and 9 G, corresponding to the first matching field $H_1$, and $H_1/2$, respectively. This measurement has been done by the same four-terminal technique used for Fig. 2, and at the same temperature $T$$=$7.07 K.  Figure 3 illustrates that the $dV/dI\ \textit{vs}.\ I_{dc}$ behavior can be divided into three regimes: I, II and III \cite{yaron}. In regime I, the measurements show zero differential resistance, which indicates that the vortex lattice is static due to the strong pinning force. In regime II, the differential resistance increases gradually when we increase the dc current. This regime is called the plastic flow regime, where the vortex lattice has been partly disordered and shows plastic flow behavior. In regime III, the measurements show a sharp resistance jump to the normal state of the film, which indicates that the vortex lattice has been completely disordered and shows avalanche motion. In our measurements, there exist in addition three sub-regimes in the plastic flow regime, which are labeled II(1), II(2) and II(3) in Fig. 3. In II(1) the measurements show the standard matching field anomaly, \textit{i.e.}, the 18 G curve, where each flux line is expected to be tied to one APC, has a smaller resistance compared with the 9 G curve. In II(2), the differential resistance measured at 18 G is larger than the one measured at 9 G, which is opposite to the results in regime II(1). In II(3), the 18 G curve and the 9 G curve merge together. No oscillations appear in this sub-regime. It should also be pointed out that the crossing point between regime II(1) and II(2) is quite sensitive to the system temperature. As shown in Fig. 3(b), this crossing point moves from 103 $\mu$A to 152 $\mu$A when we cool our system down just 0.01 K, from 7.07 K to 7.06 K. In other words, the novel behavior in the regime II(2) only happens in a very small temperature range, which might be the reason why it has not been observed in previous work.

To our knowledge, no current theory can explain this unusual behavior. We propose here a phenomenological picture to explain our experimental results. The total force \textbf{\textit{f}} acting on each flux line in a superconducting film with APCs can be expressed as $\textbf{\textit{f}}=\textbf{\textit{f}}^{vp}+\textbf{\textit{f}}^{vv}+\textbf{\textit{f}}_{T}+\textbf{\textit{f}}_{d}$, where $\textbf{\textit{f}}^{vp}$ is the pinning force due to the trapping of the flux line by a APC, $\textbf{\textit{f}}^{vv}$ is the repulsive interaction with other flux lines, $\textbf{\textit{f}}_{T}$ the force associated with thermal fluctuations, and $\textbf{\textit{f}}_{d}$ the Lorentz force, present when transport current is sent though the film \cite{reichhardt1,reichhardt2}. The first two terms in $\textbf{\textit{f}}$ would result in a highly ordered vortex lattice configuration, especially at each matching field; the last two terms drive the system from the highly ordered vortex lattice state to a disordered state. The motion of flux lines is determined by the balance between $\textbf{\textit{f}}^{vp}+\textbf{\textit{f}}^{vv}$ and $\textbf{\textit{f}}_{T}+\textbf{\textit{f}}_{d}$. In our experiments, we fix the system temperature at a value slightly below the superconducting transition temperature of the Nb film, that is, we set $\textbf{\textit{f}}_{T}$ to be constant, and vary $\textbf{\textit{f}}_{d}$ by tuning the dc current through the device.

The behavior demonstrated in Fig. 3 might be associated with a change of the flow mode of the flux lines when the dc current is increased. At very  low dc currents (regime I), the pinning force dominates the system, $(\textbf{\textit{f}}^{vp}+\textbf{\textit{f}}^{vv})$$\gg$$(\textbf{\textit{f}}_{T}+\textbf{\textit{f}}_{d})$. As a consequence, the flux lines are static and the measurements show zero differential resistance.   This corresponds to the situation shown in Fig. 4(a), which shows a schematic of the potential landscape seen by the vortex lines at $H_1$. There is a minimum of potential at each APC that pins exactly one vortex line.  Any vortex line that is displaced from its pinning center by thermal fluctuations is immediately forced back due to the repulsive interactions between vortices, leading to a minimum in the resistance due to vortex motion.  For half-filling (corresponding to $H_1/2$, the situation shown in Fig. 4(b)), the distance between the vortices is larger, leading to a smaller vortex-vortex interaction.  Each vortex is therefore not so rigidly pinned as in Fig. 4(a), giving rise to a larger resistance due to vortex motion.

The effect of applying a dc current is to add a finite slope to the potential profile, as shown in Figs. 4(c)-(f). The larger the current, the steeper the slope. At low $I_{dc}$, (corresponding to region II(1)), the pinning potential is still strong enough that the situation described above for Figs. 4(a) and (b) still holds, and the resistance at $H_1$ (corresponding to Fig. 4(c)) is less than the resistance at $H_1/2$ (Fig. 4(d)).  At still higher $I_{dc}$ (corresponding to region II(2)), however, the situation changes.  Consider the case of full filling, at $H_1$, shown schematically in Fig. 4(e).  Here, the effective pinning potential at one end of a pinning center is effectively reduced by the slope of the potential induced by the dc field, so that thermal fluctuations can easily kick a flux line out of the potential well.   If this happens in the center of the superconducting film, the repulsive interactions between vortices are still likely to push the vortex line back into its pinning site.  If the vortex pops out of a pinning site at the edge of the film, however, it will leave the film altogether.  At this point, the flux line at the pinning site immediately adjacent to the now empty site will no longer feel the repulsive interaction due to the absent vortex line.  If the repulsive interaction from its remaining vortex neighbors is sufficient to overcome the reduced pinning potential, it will jump to the empty pinning center. This jump mechanism will be continued down the linear chain of vortices, leading to a mass movement of vortices in the lattice, and hence a higher resistance.  For half-filling ($H_1/2$, shown in Fig. 4(f)), the smaller repulsive interaction between vortices, arising from the greater inter-vortex distance, is not sufficient to overcome the effective pinning potential, so that the flux lines do not move.  Hence the resistance of the sample at $H_1/2$ is lower than its resistance at $H_1$.  This mechanism clearly depends on a fine balance between the dynamic pinning potential, which depends on $I_{dc}$, and the repulsive vortex-vortex interactions, and hence can only be seen in a narrow range of $I_{dc}$.  At even higher currents, corresponding to region II(3), the potential profile is tilted so much by the applied dc current that the vortex lattice flows at both $H_1/2$ and $H_1$, so that no clear oscillations are seen in the magnetoresistance. A similar picture also holds at higher magnetic field, where there is more than one vortex per unit cell. One might also point out that vortex-antivortex pairs are present even at zero field, as evidenced by the presence of a finite resistance.

An alternative explanation is based on the remarkable similarity between the data shown in Fig. 2, and the oscillations in conductance observed in an Aharonov-Bohm interferometer formed from a quantum dot \cite{yacoby}. In that experiment, a similar change in the phase of the oscillations is observed as the phase difference between the two arms of the interferometer is varied by means of an external gate potential. Indeed, in that experiment, the oscillations at the fundamental period $\Phi_0$ disappear at the cross-over point, and only oscillations at $\Phi_0/2$ remain. Hints of this behavior are also seen in our data (see curve (d) in Fig. 2). This suggests that the transport current in our system is modulating the phase of the interference around each antidot in the lattice. The exact nature of this modulation is not clear to us at the moment, and further investigation is required to understand the origin of this unusual behavior.

This work is supported by the NSF under grant number DMR-0201530.

\newpage
%\vspace{-0.6cm}

%------------FIGURE CAPTIONS---------------------------------
%\newpage
%\begin{figure}
%\caption{(a) Schematic of the device. The entire device is made of a single layer of Nb film with a square lattice of APCs. The middle section between the two voltage leads, the part measured, is 50-$\mu$m-wide and 290-$\mu$m-long. (b) A scanning electron microscopy (SEM) image of the Nb film.}
%\label{fig1}
%\end{figure}
%\begin{figure}
%\caption{Differential resistance of the Nb film as a function of magnetic field normal to the film at different dc bias currents ($I_{dc}$) at a temperature of $T$$=$7.07 K: (a)$I_{dc}$$=$61 $\mu$A; (b)$I_{dc}$$=$70 $\mu$A; (c)$I_{dc}$$=$100 $\mu$A; (d)$I_{dc}$$=$110 $\mu$A; (e)$I_{dc}$$=$120 $\mu$A; (f)$I_{dc}$$=$140 $\mu$A; (g)$I_{dc}$$=$159 $\mu$A.}
%\label{fig2}
%\end{figure}
%\begin{figure}
%\caption{Differential resistance of the Nb film as a function of dc current at two different magnetic fields of 18 G (solid line) and 9 G (dashed line), at two different temperatures: (a) $T$$=$7.07 K and (b) $T$$=$7.06 K.}
%\label{fig3}
%\end{figure}
%\begin{figure}
%\caption{Schematics of the potential landscape seen by the vortex lines at different dc bias currents and at two different magnetic fields: $H_1$ (left column panel) and $H_1/2$ (right column panel). Each flux line is denoted by an arrow. As predicted by theory, an empty APC shows a deeper potential compared with one filled by a flux line \cite{mkrtchyan}.}
%\label{fig4}
%\end{figure}
%------------FIGURES---------------------------------
\newpage
\center{Fig. 1, Z. Jiang \etal}
\vspace{3cm}
\begin{figure}[h]
\includegraphics[width=8cm]{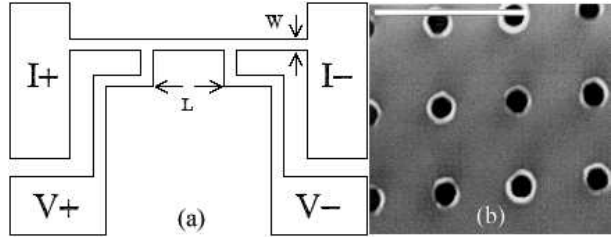}
\caption{(a) Schematic of the device. The entire device is made of a single layer of Nb film with a square lattice of APCs. The middle section between the two voltage leads, the part measured, is 50-$\mu$m-wide and 290-$\mu$m-long. (b) A scanning electron microscopy (SEM) image of the Nb film.}
\label{fig1}
\end{figure}

\newpage
\center{Fig. 2, Z. Jiang \etal}
\vspace{3cm}
\begin{figure}[h]
\includegraphics[width=8cm]{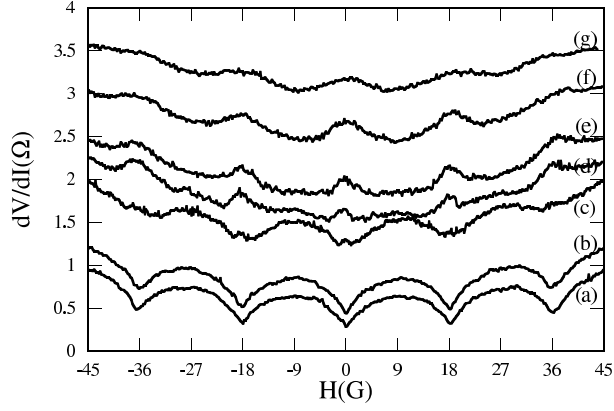}
\caption{Differential resistance of the Nb film as a function of magnetic field normal to the film at different dc bias currents ($I_{dc}$) at a temperature of $T$$=$7.07 K: (a)$I_{dc}$$=$61 $\mu$A; (b)$I_{dc}$$=$70 $\mu$A; (c)$I_{dc}$$=$100 $\mu$A; (d)$I_{dc}$$=$110 $\mu$A; (e)$I_{dc}$$=$120 $\mu$A; (f)$I_{dc}$$=$140 $\mu$A; (g)$I_{dc}$$=$159 $\mu$A.}
\label{fig2}
\end{figure}

\newpage
\center{Fig. 3, Z. Jiang \etal}
\vspace{3cm}
\begin{figure}[h]
\includegraphics[width=8cm]{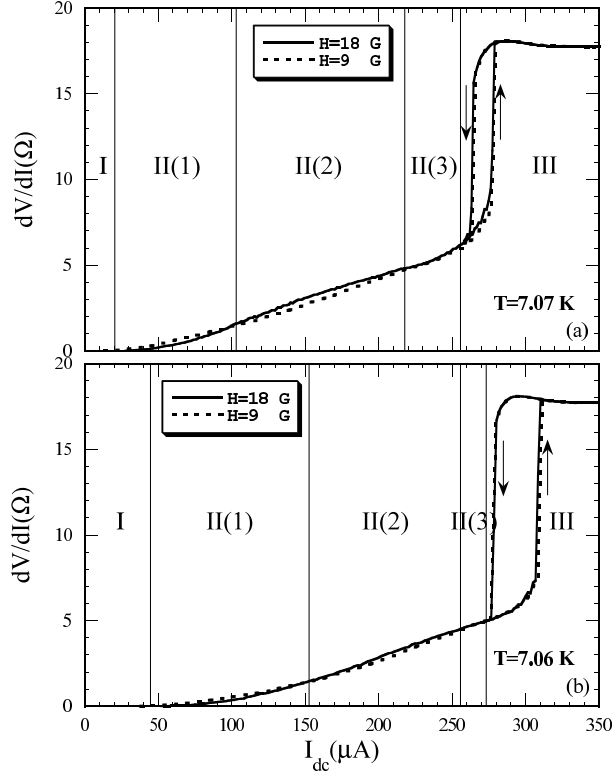}
\caption{Differential resistance of the Nb film as a function of dc current at two different magnetic fields of 18 G (solid line) and 9 G (dashed line), at two different temperatures: (a) $T$$=$7.07 K and (b) $T$$=$7.06 K.}
\label{fig3}
\end{figure}

\newpage
\center{Fig. 4, Z. Jiang \etal}
\vspace{3cm}
\begin{figure}[h]
\includegraphics[width=8cm]{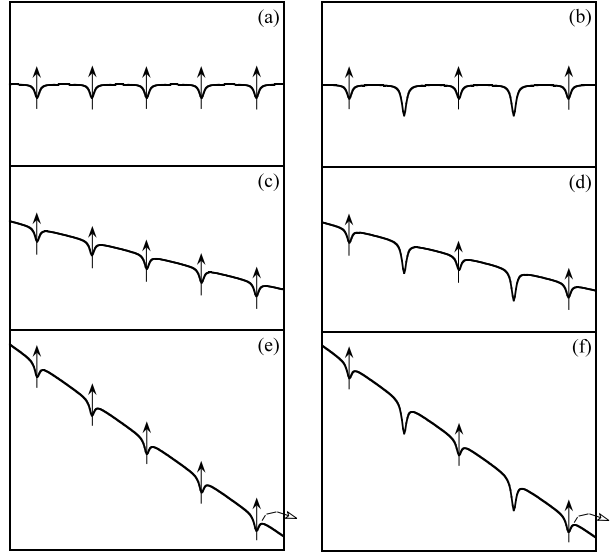}
\caption{Schematics of the potential landscape seen by the vortex lines at different dc bias currents and at two different magnetic fields: $H_1$ (left column panel) and $H_1/2$ (right column panel). Each flux line is denoted by an arrow. As predicted by theory, an empty APC shows a deeper potential compared with one filled by a flux line \cite{mkrtchyan}.}
\label{fig4}
\end{figure}

\end{document}